# JOHN G. CRAMER

## The Plane of the Present and the New Transactional Paradigm of Time

### 1. Introduction.
### Time and The Plane of the Present
### as Evolving Paradigms

The *plane of the present* is a concept that is useful for discussing the various paradigms of time. Here by 'plane of the present' we mean the temporal interface that represents the present instant and that forms the boundary between the past and the future. We use the geometrical term 'plane' to indicate an extended surface in the space-time continuum, as opposed to a 'point' on some time axis. This point/plane dichotomy is intended to raise issues of extension and simultaneity and to examine the degree to which these are meaningful concepts from various physical viewpoints. We will show by example in the present work that the plane of the present is a pivotal concept that offers considerable power in differentiating between various views of the nature of time.

The concept of time within the main stream of physics thinking has followed a rather convoluted path over the past three millennia. Anticipating the modern motion picture, Zeno of Elea (c.490-c.430 B.C) questioned whether time should appropriately be viewed as a continuously flowing river, or should more properly be considered as a rapid sequence of stop-motion 'freeze-frames', in effect rendering geometrical each instant as a separate infinitesimal point on the line of time. Adopting this view, he asked how physical motion could occur. He argued paradoxically that motion is not possible, since it appears to happen only *between* the frozen frames of time instants.[1]

From the viewpoint of Zeno, the plane of the present would be simply the last and most recent in this sequence of freeze-frames. It would be that frozen instant, spanning the universe, which changes progressively as the instant we call 'now' becomes the frozen past and future possibility freezes into the 'now' of present reality. We note that the plane of the present as a concept does not resolve the arrow paradox that Zeno raised. It only provides a way of thinking about it.

Isaac Newton (1642 - 1727) changed the paradigm of time by introducing the concept of absolute time, a sequence of instants that are the same at all points in space. Thus the present is a locus of a universal time characterizing a three dimensional volume of space at some instant during its dynamic evolution. For Newton, time was an absolute clock that ticked everywhere at the same rate. He altered and extended Zeno's instant of time by inventing differential calculus, thereby transforming the frozen time frames to *time derivatives*, the essential embodiment of change in the infinitesimal limiting case. In the process, Newton implicitly provided the solution to Zeno's arrow paradox by providing a formalism in which motion and change are explicitly a part of each freeze frame.

The mathematical concept of a series expansion around an instant of time, a later extension of Newton's ideas, implies that the entire past and future of any continuous functional behavior is embodied in the values of the time-dependent function and all of its derivatives, evaluated only at one single instant of time. This led to the $19^{th}$ century paradigm of a 'clockwork universe' in which, given the positions and velocities of all particles at some instant of time and a knowledge of their interaction forces, the entire past and future of the universe was determined and, in principle, could be calculated.

In such a deterministic universe, the plane of the present loses much of its significance. The present is simply the location of the 'bead' representing our consciousness as it slides along a fixed wire that spans the frozen past and the pre-determined future.

Albert Einstein (1879 - 1955) changed the time paradigm yet again, rejecting Newton's absolute universal time and instead depicting time as mutable, a fourth dimension that in some sense is interchangeable with the three spatial dimensions. Each inertial reference frame (i.e., observer moving with some constant velocity), in Einstein's view, achieves its own particular trade-off between the local space and time dimensions. In this democracy of dimensions, time is placed on a more equal footing with space. This led to the concept of a 'block universe', an infinite stack of three dimensional infinitesimal time-slices of the universe, in which all regions of time, future as well as past, must co-exist in the overall four-space continuum. Each observer-based reference frame makes its own angled time-slice through this block, representing the view of that observer as to which event-points in the four-dimensional block occur at the same time.

With special relativity Einstein destroyed the Newtonian concept of absolute time. In Einstein's world it is not meaningful to speak of two spatially separated events as truly 'simultaneous'. Instead, the relative time of occurrence of separated space-time events depends on the reference frame of the observer. The reference frame also affects the rate at which time progresses, so that time is slowed in a moving reference frame, with the question of which frame is moving also depending on the observer.

The plane of the present in the Einsteinian view is thus not an immutable surface that always cuts through space-time in the same way, but rather a mutable plane that may tilt at an angle that depends on the observer's reference frame. Moreover, Einstein showed, with the introduction of general relativity, that time slows, and indeed may stop altogether, in a sufficiently strong gravitational field. Therefore, time may run at different rates in locations with different strengths of the local gravitational field. The concept of 'proper time' in a reference frame is all that remains of Newtonian time. Some plane of the present (or surface of simultaneity) may connect events in different regions of this time-distorted space-time, but it has become an elusive and convoluted concept.

Werner Heisenberg (1901 - 1976) again changed the time paradigm with his formulation of quantum mechanics, the physics theory of matter and energy at the smallest scales.[2] Heisenberg found that the *uncertainty principle* was contained within his new quantum formalism.[3] The uncertainty principle is in essence a new physical law showing that there is a certain minimum uncertainty between certain pairs of 'complementary' physical variables. These uncertain pairs of complementary variables include position and momentum (along any coordinate axis) and (of particular relevance for the present discussion) time and energy.

The uncertainty principle demonstrates that the concept of the instant is a geometrical ideal rather than a physical concept, because the uncertainties in the determinations of time and energy are inextricably entangled within the time-energy uncertainty relation. Any measurement of energy and/or any constraints on energy in a physical process will unavoidably bring with it an uncertainty in the time variable. Consequently, since some information about energy is always available in any real physical process, the concept of an infinitesimal instant of time is a geometrical idealization that cannot be applied to the physical universe. Similarly, a precise measurement of velocity (or momentum, to be precise) brings with it an unavoidable uncertainty in position, so the concept of a mathematical point representing position is also inappropriate.

The Newtonian clockwork universe is thus destroyed by the uncertainty principle. Heisenberg's indeterminacy means that the past and future are not determinable from the data of an instant. The future, outside a very limited region, is uncertain and unpredictable.

Correspondingly, the plane of the present, instead of being the perfectly sharp edge implied by the Newtonian view, is a boundary that is blurred by the uncertainty principle. In the quantum world, the plane of the present must be a fuzzy region with the past on one side in time and the future on the other, but with some uncertainty in the central region about which is the past and which is the future. Across this region there must occur the freezing of possibility into reality, but possibility and actuality are mixed and smeared by time indeterminacy.

As we have seen from the preceding discussion, the evolution of the paradigm of time was driven not by the speculations of philosophers but by physics, the activity of checking concepts against the structure of the universe. This process has periodically required revision of our ideas, as our ability to do such checking has improved. This is an ongoing process, the next step of which may come from an interpretation of quantum mechanics.

## 2. The Transactional Interpretation of Quantum Mechanics

Quantum mechanics is unique among the major theories of physics in having emerged from the mathematical formulations of Schrödinger and Heisenberg in the 1920s without being accompanied by a clear picture of the physical processes described by the mathematical formalism. The mathematical formulation of quantum mechanics is now well established and tested, but there is still great controversy concerning the *interpretation* of that formalism[4].

In the formalism of quantum wave mechanics, a physical system is represented by a second-order differential equation or 'wave equation'. Maxwell's electromagnetic wave equation and the Schrödinger equation are two examples of wave equations used in quantum mechanics. Within the wave equation, the physical properties of a system are represented by boundary conditions that characterize known limits and regions of applicability and by a potential that characterizes the way in which energy changes when elements of the system are moved or displaced. The wave equation is solved mathematically to produce a solution called a 'wave function'. The wave function is then used, applying standard quantum mathematical procedures, to make predictions about the probable outcomes (or 'expectation values') of physical measurements that might be made on the system, e.g., its position, momentum, energy, etc.

In the 1920s when quantum mechanics was introduced, there was perceived to be a serious problem. The two major physical theories developed in the first decades of the 20$^{th}$ century, special relativity and quantum mechanics, were initially viewed as incompatible. This incompatibility arose because the Schrödinger equation, the wave equation used to describe the behavior of particles of non-zero rest mass, is inconsistent with relativity. Fortunately, later developments in quantum mechanics produced alternative wave equations, the Klein-Gordon equation for integer-spin bosons and the Dirac equation for half-integer spin fermions, which are fully consistent with relativity. Further, it was found that the Schrödinger equation could be viewed as the non-relativistic (low-velocity) limit of either of these more correct wave equations[5].

While the wave function has proved to be an extremely useful mathematical object for applications of quantum mechanics, its *meaning* remains a matter of heated debate. In classical mechanics, such a wave function would represent a wave physically present in space (for example, a traveling electromagnetic radio wave), but in quantum mechanics this simple explanation of the wave function is conventionally rejected. Since the late 1920s the orthodox view, as embodied in the Copenhagen interpretation of quantum mechanics, has been that a wave function is *not* physically present in space, but rather is a purely mathematical construct used for calculation, which can be interpreted as a mathematical function encoding the state of knowledge of some observer. While the Copenhagen view is self-consistent, it prevents the visualization of quantum mechanical processes. Further, careful considerations of the implications of the Copenhagen interpretation have generated a number of 'interpretational paradoxes',[6] e.g., Schrödinger's Cat, Wigner's Friend, Wheeler's Delayed Choice, the Einstein-Podolsky-Rosen paradox, and so on, that lead into very deep philosophical waters.

The *transactional interpretation of quantum mechanics*,[7] originally presented in a long review article in 1986, is an alternative to the Copenhagen interpretation that avoids these paradoxes. It is relativistically invariant, so that it can be used with the relativistic wave equations discussed above, and it uses the advanced and retarded wave function solutions of these equations in a 'handshake' that provides a rationale for treating the wave function as physically present in space.

The logical development of the transactional interpretation starts with the time-symmetric classical electromagnetism of Dirac[8], and Wheeler and Feynman[9] which describes electromagnetic processes as an exchange between retarded (normal) and advanced (time-reversed) electromagnetic waves. The transactional interpretation applies the time-symmetric Wheeler-Feynman view to the quantum mechanical wave function solutions of the electromagnetic wave equation. The lessons learned about electromagnetic quantum waves are then extended to wave functions describing the behavior of massive particles (e.g., electrons and protons) by applying the same interpretation to their relativistic wave equations. Finally, the Schrödinger equation is included as a non-relativistic reduction of the relativistic wave equations in the limit of small velocities.

The transactional interpretation views each quantum event as a 'handshake' or 'transaction' process extending across space-time that involves the exchange of advanced and retarded waves to enforce the conservation of certain quantities (energy, momentum, angular momentum, etc.). It asserts that each quantum transition forms in four stages: (1) *emission*, (2) *response*, (3) *stochastic choice*, and (4) *repetition to completion*.

The first stage of a quantum event is the *emission* of an 'offer wave' by the 'source', which is the object supplying the quantities transferred. The offer wave is the time-dependent retarded quantum wave function $\Psi$, as used in standard quantum mechanics. It spreads through space-time until it encounters the 'absorber', the object receiving the conserved quantities.

The second stage of a quantum event is the *response* to the offer wave by any potential absorber (there may be many in a given event). Such an absorber produces an advanced 'confirmation wave' $\Psi^*$, the complex conjugate of the quantum offer wave function $\Psi$. The confirmation wave travels in the reverse time direction and arrives back to the source at precisely the instant of emission with an amplitude of $\Psi\Psi^*$.

The third stage of a quantum event is the *stochastic choice* exercised by the source in selecting one from among the possible transactions. It does this in a linear probabilistic way based on the strengths $\Psi\Psi^*$ of the advanced-wave 'echoes' it receives from the potential absorbers.

The final stage of a quantum event is the *repetition to completion* of this process by the source and absorber, reinforcing the selected transaction repeatedly until the conserved quantities are transferred and the potential quantum event becomes real.

Since the advanced-retarded-wave handshake used by the transactional interpretation operates in both time directions, it is in a sense atemporal, in that no elapsed time at the space-time site of the emitter is required between the beginning of a quantum event as an offer wave and its conclusion as a completed transaction (or 'collapsed wave function', in the terminology of the Copenhagen interpretation). Similarly, at the space-time site of the absorber (the future end of a transaction), there is no elapsed time between responding with a confirmation wave and the completion of the transaction. The transactional interpretation asserts that at the quantum level time is a two-way street, in which at some level the future determines the past as well as the past determining the future.

However, there is time direction preference associated with quantum mechanics. In the formalism of quantum mechanics, processes are usually analyzed using what is called the 'post' formalism. In terms of the transactional interpretation, this is the assumption that the probability of a quantum transition $\Psi\Psi*$ is evaluated at the space-time location of the emitter (the 'past' end of the transaction) rather than at the location of the absorber (the 'future' end of the transaction). The alternative is the time-reversed

'prior' formalism, in which the probability is evaluated at the future absorber end of the process.[10] Normally in quantum calculations the two formalisms, if calculated with sufficient accuracy, give the same result, but there are cases involving violations of time reversal invariance where they do not. Therefore, the transactional interpretation, despite its even-handed treatment of advanced and retarded waves, implies a limited preference for a time direction on the basis of its description of transaction formation.

### 3. The Hierarchy of the Arrows of Time

In the situation described above in which quantum wave functions traveling in both time directions are handled in an even-handed and symmetric way, one must ask about the origins of the macroscopic 'arrow of time' that is evident in the everyday world. This question is complicated by the presence of at least five seemingly independent 'arrows of time' that can be identified in the physical world. Let me briefly review them.

1. *Subjective:* At the macroscopic level, it is self-evident that the past and the future are not the same. We remember the past but not the future. Our actions and decisions can affect the future but not the past.
2. *Electromagnetic:* We can send electromagnetic signals to the future but not to the past. A current through an antenna makes retarded positive-energy waves but not advanced negative-energy waves.
3. *Thermodynamic:* Isolated systems have low entropy (i.e., disorder) in the past and gain entropy and become more disordered in the future. Molecules released from a confining box will rapidly fill a larger volume, but they will not spontaneously collect themselves back in the box.
4. *Cosmological:* The universe was smaller and hotter in the past but will be larger and cooler in the future. The time direction of expansion is an arrow.
5. *CP Violation:* The $K°_L$ meson (the neutral long-lived K meson, which is a matter-antimatter combination of a down quark and a strange quark) exhibits weak decay modes having matrix elements and transition probabilities that are larger for the decay process than for the equivalent time-reversed process, in violation of the principle of time-reversal invariance. This is related to the so-called CP violation[11] that is present in the neutral K meson system, which shows a preference for matter over antimatter in certain decay processes.

These indications of time asymmetry cannot be independent, and one would like to understand their connections and hierarchy. It is generally agreed that the CP violation arrow is probably the most fundamental time arrow and is responsible for both the dominance of matter in the universe and the breaking of time symmetry in fundamental interactions to make possible the expansion of the universe in a particular time direction, thereby leading to the cosmological arrow of time. From this point on, however, there are at least two divergent views of the hierarchy.

    The first of these views might be characterized as the orthodox view of hierarchy, in that it has been widely advocated in the physics literature by many authors, particularly Hawking.[12] It is illustrated by Figure 1(a) and asserts that the thermodynamic arrow follows from the cosmological arrow and leads to the electromagnetic and subjective arrows of time.

This view goes back to Ludwig Boltzmann (1844-1906), who in the 1870s 'derived' from first principles the thermodynamic arrow of time and the 2nd law of thermodynamics with his famous H-Theorem. If, as Boltzmann appears to demonstrate, the increase in entropy can be derived from first principles using only statistical arguments, this would provide strong support for its primary position in the hierarchy of time arrows.

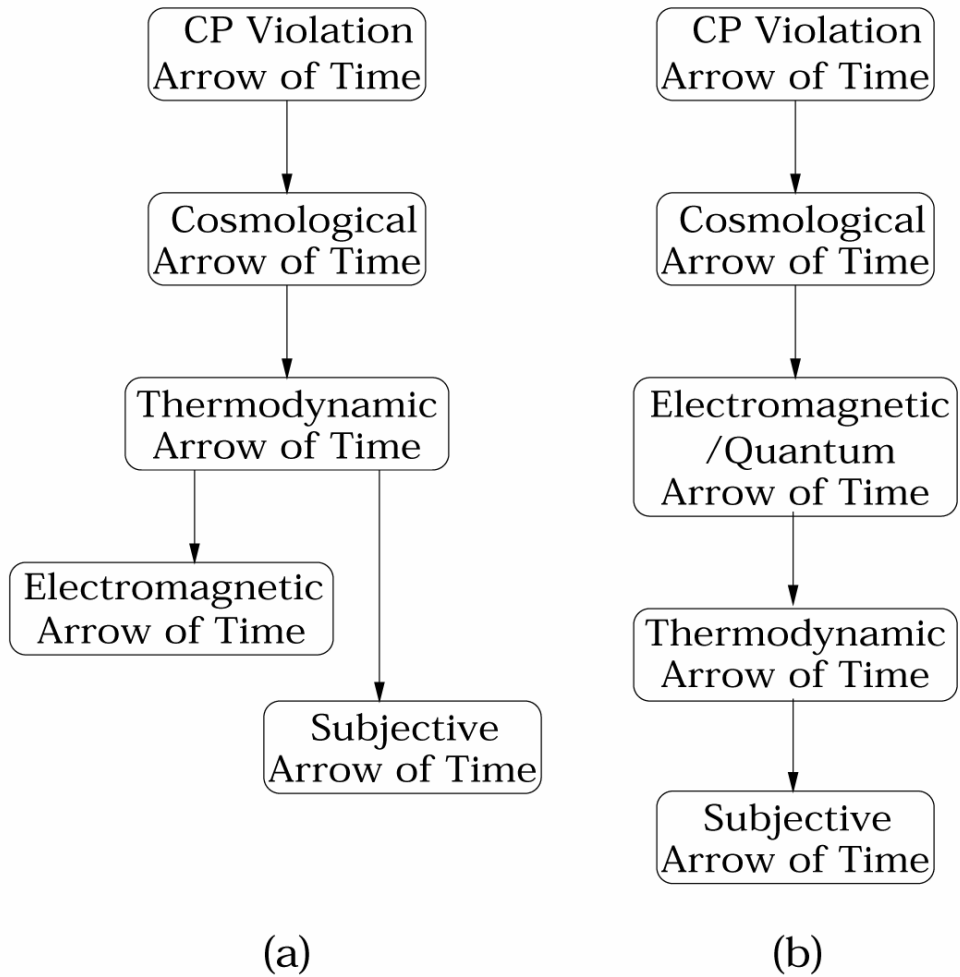

(a)        (b)

**Figure 1. Two alternative hierarchies of the arrows of time: (a) the hierarchy generally favored in the physics literature; and (b) the hierarchy advocated here.**

However, the 'first principles' that lead to Boltzmann's H-Theorem include an implicit time-arrow. Boltzmann used the apparently reasonable assumption that the motions of colliding members of a system of particles are uncorrelated *before a collision*. Unfortunately, that assumption is not as innocent as it appears. It smuggles into the problem an implicit time asymmetry, which ultimately leads to a system entropy that is constant or increasing with time. However, if one, in the spirit of extending the H-theorem, had assumed that the motions of colliding particles were uncorrelated *after the collision*, then one would have demonstrated with equal rigor that the entropy was *constant or decreasing* with time, i.e., the thermodynamic arrow would be pointing in the wrong direction.

This consideration leads to the view of the time arrow hierarchy illustrated by Figure 1(b), which asserts that the electromagnetic arrow of time produces the thermodynamic arrow because of the dominance of retarded electromagnetic waves and interactions in the universe. The source of the time asymmetry implicit in the assumptions of the H-Theorem is the intrinsic 'retarded' character of electromagnetic interactions. Here the term retarded means that there is a speed-of-light time delay between the occurrence of some change in a source of electromagnetic field, i.e., the movement of an electric charge, and its appearance in distant electromagnetic fields produced by that source. The field change always occurs *after* the source change, in any reference frame.[13]

The time arrows listed above did not include a quantum mechanical arrow of time. However, we have observed that from the transactional viewpoint the present source or emitter selects from among the possible quantum transaction, not the future absorber. How does this time preference fit into the above hierarchy?

In effect, this quantum arrow of time is equivalent to the electromagnetic arrow, which requires the dominance of retarded waves. The conditions of our universe favor retarded waves and positive energies. While quantum processes may include the actions of advanced waves to enforce conservation laws, no net 'advanced effects' are allowed. Such effects would represent violations of the principle of causality, and have not been observed. We conclude that the hierarchy indicated in Figure 1(b), the primacy of the electromagnetic/quantum arrow, is consistent with the quantum time preference discussed above and can be used along with a more enlightened version of Boltzmann's H-Theorem to derive the thermodynamic arrow of time.

## 4. Determinism and the Transactional Interpretation

It has been asserted[14] that the transactional interpretation is necessarily deterministic, requiring an Einsteinian block universe to pre-exist, because the future must be fixed in order to exert its influence on the past in a transactional handshake. However, while block-universe determinism is consistent with the transactional interpretation, it is not required. A part of the future is emerging into a fixed local existence with each transaction, but the future is not determining the past, and the two are not locked together in a rigid embrace.

Let us make an analogy. The handshakes envisioned by the transactional interpretation bear some resemblance to the handshakes that take place on telephone or Internet lines these days when one uses a debit card to make a purchase at a shop. The shop's computer system reads the magnetic strip on your card and transmits the information to the bank, which verifies that your card is valid and that your bank balance is sufficient for the purchase, and then removes the purchase amount from your bank balance. This transaction enforces 'conservation of money'. There is a one to one correspondence between the amount the store receives for your purchase and the amount that is deducted from your bank account. The transaction assures that the amount is deducted only once, from only one bank account and credited only once, to only one store. On the other hand, the bank does not exert any influence on what you choose to purchase, beyond insuring that money is conserved in the transaction and that you do not overspend your resources.

A quantum event as described by the transactional interpretation follows the same kind of protocol. There is a one to one correspondence between the energy and other conserved quantities (momentum, angular momentum, spin projections, etc.) that are conveyed from the emitter to the absorber, but aside from the enforcement of conservation laws, the future absorber does not influence the past emission event. Therefore, the 'determinism' implied by the transactional interpretation is very limited in its nature.[15]

## 5. The Plane of the Present and the Transactional Interpretation of Quantum Mechanics

This brings us to the question of how our conceptualization of the plane of the present is affected by the transactional interpretation. In the transactional interpretation, the freezing of possibility into reality as the future becomes the present is not a plane at all, but a fractal-like[16] surface that stitches back and forth between past and present, between present and future.

To make another analogy, the emergence of the unique present from the future of multiple possibilities, in the view of the transactional interpretation, is rather like the progressive formation of frost crystals on a cold windowpane. As the frost pattern expands, there is no clear freeze-line, but rather a moving boundary, with fingers of frost reaching out well beyond the general trend, until ultimately the whole window pane is frozen into a fixed pattern. In the same way, the emergence of the present involves a lacework of connections with the future and the past, insuring that the conservation laws are respected and the balances of energy and momentum are preserved.

Is free will possible in such a system? I believe that it is. Freedom of choice does not include the freedom to choose to violate physical laws. The transactional handshakes between present and future are acting to enforce physical laws, and they restrict the choices between future possibilities only to that extent.

Therefore, we conclude that the transactional interpretation does not require a deterministic block universe. It does, however, imply that the emergence of present reality from future possibility is a far more complex process than we have previously been able to imagine. This is the new transactional paradigm of time.

Department of Physics
University of Washington
Seattle, Washington, USA

# Notes

[1] We note that modern quantum optics has provided a new demonstration of Zeno's viewpoint, with the freeze-frames taking the form of the collapse of the quantum wave-function following measurement and the continuous time evolution occurring with the absence of measurements (see P.G. Kwiat, et al., 'High-efficiency Quantum Interrogation Measurements via the Quantum Zeno Effect', *Phys. Rev. Letters* **83**, (1999) 4725-4728; also available in the Internet at http://xxx.lanl.gov/abs/quant-ph/9909083). Because a quantum measurement of a time-evolving system can "reset" the system to its initial state repeatedly, thereby blocking the time evolution that would otherwise occur, the repeated measurements can freeze Zeno's arrow in one particular freeze-frame and prevent its motion.

[2] W. Heisenberg, 'On the Quantum Interpretation of Kinematical and Mechanical Relationships', *Zeitscrift für Physik* **33** (1925), 879-893.

[3] W. Heisenberg, 'The Actual Content of Quantum Theoretical Kinematics and Mechanics', *Zeitscrift für Physik* **43** (1927), 172-198.

[4] There are actually two quantum formalisms, the matrix mechanics of Heisenberg and the wave mechanics formalism of Schrödinger and Dirac. However, since these have been shown to be completely equivalent, we will focus on Schrödinger-Dirac wave mechanics.

[5] We note that quantum field theory, the most widely used formalism in high energy physics, is consistent with special relativity and avoids wave equations altogether, while the task of making quantum mechanics compatible with *general relativity*, the current standard model of gravity, remains an unsolved problem.

[6] J.G. Cramer, 'The Transactional Interpretation of Quantum Mechanics', *Reviews of Modern Physics* **58** (1986), 647-687; also available on the Internet at http://faculty.washington.edu/jcramer/theory.html.

[7] J.G. Cramer, 'The Transactional Interpretation of Quantum Mechanics'; J.G. Cramer, 'An Overview of the Transactional Interpretation of Quantum Mechanics', *International Journal of Theoretical Physics* **27** (1988), 227-236; also available on the Internet at http://faculty.washington.edu/jcramer/theory.html.

[8] P.A.M. Dirac, 'Classical Theory of Radiating Electrons', *Proc. Royal Soc. London* **167A** (1938), 148-169.

[9] J.A. Wheeler & R.P. Feynman, 'Interaction with the Absorber as the Mechanism of Radiation', *Reviews of Modern Physics* **17 (1945)**, 157-181; J.A. Wheeler & R.P. Feynman, 'Classical Electrodynamics in Terms of Direct Interparticle Action', *Reviews of Modern Physics* **21** (1949), 425-434.

[10] R.M. DeVries, G.R. Satchler, and J.G. Cramer, 'Importance of Coulomb Interaction Potentials in Heavy-Ion Distorted Wave Born Approximation Calculations', *Phys. Rev. Letters* **32** (1974), 1377.

[11] In quantum mechanics, CP is a symmetry transformation involving the simultaneous conversion of matter particles to antimatter and vice versa (C) and the reversal of the three spatial coordinate axes (P).

[12] Stephen W. Hawking, 'Arrow of Time in Cosmology', *Phys. Rev.* **D32** (1985), 2489.

[13] For further discussion of the arrow of time problem, see J.G. Cramer, 'The Arrow of Electromagnetic Time and Generalized Absorber Theory', *Foundations of Physics* **13** (1983), 887-902; also available on the Internet at http://faculty.washington.edu/jcramer/theory.html; and J.G. Cramer, 'Velocity Reversal and the Arrows of Time', *Foundations of Physics* **18** (1988), 1205-1212; also available on the Internet at http://faculty.washington.edu/jcramer/theory.html.

[14] J.F. Woodward, 'Making the Universe Safe for Historians; Time Travel and the Laws of Physics', *Foundations of Physics Letters* **8** (1995), 1-40.

[15] The same kind of limited connections and information transfer, for similar reasons, is present in so-called EPR experiments involving measurements on entangled quantum subsystems. The non-local quantum connection between a pair of separated measurements preserves correlations and enforces conservation laws without permitting the transfer of information from observer to observer.

[16] The time-interface surface cannot be truly fractal in its structure because the scale terminates at the quantum limit. A quantum system cannot be 'turtles all the way down'.